\documentclass[%
 showpacs,
 showkeys,
 amsmath,amssymb,
]{revtex4-1}

\usepackage{graphicx}
\usepackage{dcolumn}
\usepackage{bm}

\usepackage{CJK}

\usepackage{color}
\usepackage{amssymb}
\usepackage{amsfonts}

\usepackage[colorlinks,urlcolor=blue,linkcolor=blue,citecolor=blue,anchorcolor=blue]{hyperref}

\begin{document}


\preprint{APS/123-QED}

\title{Nonparaxial accelerating Talbot effect}

\author{Yiqi Zhang$^1$}
\email{zhangyiqi@mail.xjtu.edu.cn}
\author{Hua Zhong$^1$}
\author{Milivoj R. Beli\'c$^{2}$}
\author{Changbiao Li$^1$}
\author{Zhaoyang Zhang$^1$}
\author{Feng Wen$^1$}
\author{Yanpeng Zhang$^{1}$}
\author{Min Xiao$^{4,5}$}
\affiliation{%
 $^1$Key Laboratory for Physical Electronics and Devices of the Ministry of Education \& Shaanxi Key Lab of Information Photonic Technique,
Xi'an Jiaotong University, Xi'an 710049, China \\
$^2$Science Program, Texas A\&M University at Qatar, P.O. Box 23874 Doha, Qatar \\
$^3$Department of Physics, University of Arkansas, Fayetteville, Arkansas 72701, USA \\
$^4$National Laboratory of Solid State Microstructures and School of Physics, Nanjing University, Nanjing 210093, China
}%

\date{\today}

\begin{abstract}
  \noindent
We demonstrate the fractional Talbot effect of nonpraxial accelerating beams, theoretically and numerically.
It is based on the interference of nonparaxial accelerating solutions of the Helmholtz equation in two dimensions.
The effect originates from the interfering lobes of a superposition of the solutions that accelerate along concentric semicircular trajectories with different radii. Talbot images form along certain central angles, which are referred to as the Talbot angles. The fractional nonparaxial Talbot effect is obtained by choosing the coefficients of beam components properly.
A single nonparaxial accelerating beam possesses duality --- it can be viewed as a Talbot effect of itself with an infinite or zero Talbot angle.
These results improve the understanding of nonparaxial accelerating beams and the Talbot effect among them.
\end{abstract}

\maketitle

%
\section{Introduction}

Paraxial and nonparaxial accelerating beams that are based on Airy \cite{siviloglou.ol.32.979.2007,siviloglou.prl.99.213901.2007,novitsky.ol.34.3430.2009,carretero.oe.17.22432.2009},
Bessel \cite{kaminer.prl.108.163901.2012,alonso.ol.37.5175.2012},
Mathieu \cite{zhang.prl.109.193901.2012,aleahmad.prl.109.203902.2012}, and Weber wave functions \cite{zhang.prl.109.193901.2012,bandres.njp.15.013054.2013},
have attracted a lot of attention in the past decade.
They produced a variety of potential applications in particle manipulation \cite{baumgartl.np.2.675.2008,zhang.ol.36.2883.2011,schley.nc.5.5189.2014},
electron beam shaping \cite{voloch.nature.494.331.2013}, super-resolution imaging \cite{jia.np.8.302.2014}, and
surface plasmon generation \cite{zhang.ol.36.3191.2011,minovich.prl.107.116802.2011,li.prl.107.126804.2011,libster-hershko.prl.113.123902.2014},
to name a few.
Investigations of accelerating beams even have opened a new window in the exploration of elusive problems in
general relativity \cite{bekenstein.prx.4.011038.2014,bekenstein.np.11.872.2015,sheng.nc.7.10747.2016} and quantum particle physics \cite{kaminer.np.11.261.2015}.
Indeed, the development of accelerating beams is growing explosively,
and one of the most exciting scenarios is the mutual promotion and complementation between paraxial and nonparaxial accelerating beams.
One example of that promotion  is the theoretical development of nonlinear nonparaxial accelerating beams \cite{kaminer.oe.20.18827.2012}
and their experimental observation \cite{zhang.ol.37.2820.2012},
based on the linear and nonlinear paraxial accelerating beams \cite{kaminer.prl.106.213903.2011,dolev.prl.108.113903.2012,zhang.ol.38.4585.2013,zhang.oe.22.7160.2014}.
In this context, it is worth mentioning a recent investigation of incoherent paraxial and nonparaxial accelerating beams \cite{lumer.optica.2.886.2015,lumer.prl.115.013901.2015};
this work filled a gap in the understanding of accelerating beams and
helped in the management of challenging issues connected with the incoherent accelerating beams.

Not related to incoherent accelerating beams, the interference of superposed coherent accelerating beams still produces interesting results.
A new member of the Talbot effect \cite{wen.aop.5.83.2013} family, the Airy-Talbot effect, was recently introduced \cite{lumer.prl.115.013901.2015,zhang.ol.40.5742.2015}.
Different from the traditional Talbot effect \cite{iwanow.prl.95.053902.2005,zhang.prl.104.183901.2010,zhang.ieee.4.2057.2012,zhang.pre.91.032916.2015},
the accelerating Talbot effect is not based on a periodic incident beam,
but on the interference of a superposition of coherent Airy beams with transverse displacements.
The appearance of accelerating Airy-Talbot effect refreshed the understanding of the recurrence of images.
Even though both the paraxial and the nonparaxial accelerating Talbot effects were reported in \cite{lumer.prl.115.013901.2015},
the fractional nonparaxial accelerating Talbot effect was not discussed in the literature, to the best of our knowledge.
This is accomplished in this Letter.
In addition to the demonstration of fractional nonparaxial accelerating Talbot effect,
we also point out that the nonparaxial accelerating beam that accelerate along the circular trajectory has duality ---
it is a Talbot effect of itself with the Talbot angle being $\pi$ or zero.

Thus, in this Letter we establish the fractional nonparaxial accelerating Talbot effect,
by superposing nonparaxial accelerating beams with proper coefficients that accelerate along concentric semicircular trajectories.
In this investigation, we are inspired by the content of
the last paragraph in \cite{lumer.prl.115.013901.2015}.

\section{Results and Discussions}
In vacuum, the two-dimensional Helmholtz equation can be written as
\[
  \left( \frac{\partial^2}{\partial x^2} + \frac{\partial^2}{\partial z^2} \right) \vec{E} + k^2 \vec{E}=0,
\]
where $k$ is the wavenumber.
For the transverse-electric field $\vec{E}=E_y(x,z)\hat{y}$,
a particular shape-preserving solution can be written as
\begin{equation}\label{eq1}
  E_y(x,z)=\int_{-k}^k \frac{1}{k_z} \exp\left[ im \sin^{-1}\left(\frac{k_x}{k}\right)\right] \exp(ik_x x+ik_z z) dk_x,
\end{equation}
where $k=\sqrt{k_x^2+k_z^2}$ and $m$ is a real parameter that determines the radius $(\sim m/k)$ of the main lobe of the solution \cite{lumer.optica.2.886.2015}.
The solution in Eq. (\ref{eq1}) is the nonparaxial accelerating solution,
which exhibits a semicircular trajectory \cite{kaminer.prl.108.163901.2012}.
Since the solution of the Helmholtz equation is dependent on $m$, one can obtain a general solution for it, as
\begin{align}\label{eq2}
  E_y(x,z) = \int_{-k}^k dk_x  & \Bigg\{ \frac{1}{k_z} \exp(ik_x x + ik_z z) \Bigg. \times\notag \\
  & \left. \sum_{m} c_m \exp\left[ im \sin^{-1}\left(\frac{k_x}{k}\right)\right] \right\},
\end{align}
where $c_m$ is an arbitrary amplitude coefficient.
This solution is a superposition of a number of components given by Eq. (\ref{eq1}).

\begin{figure}[htbp!]
  \centering
  \includegraphics[width=0.5\columnwidth]{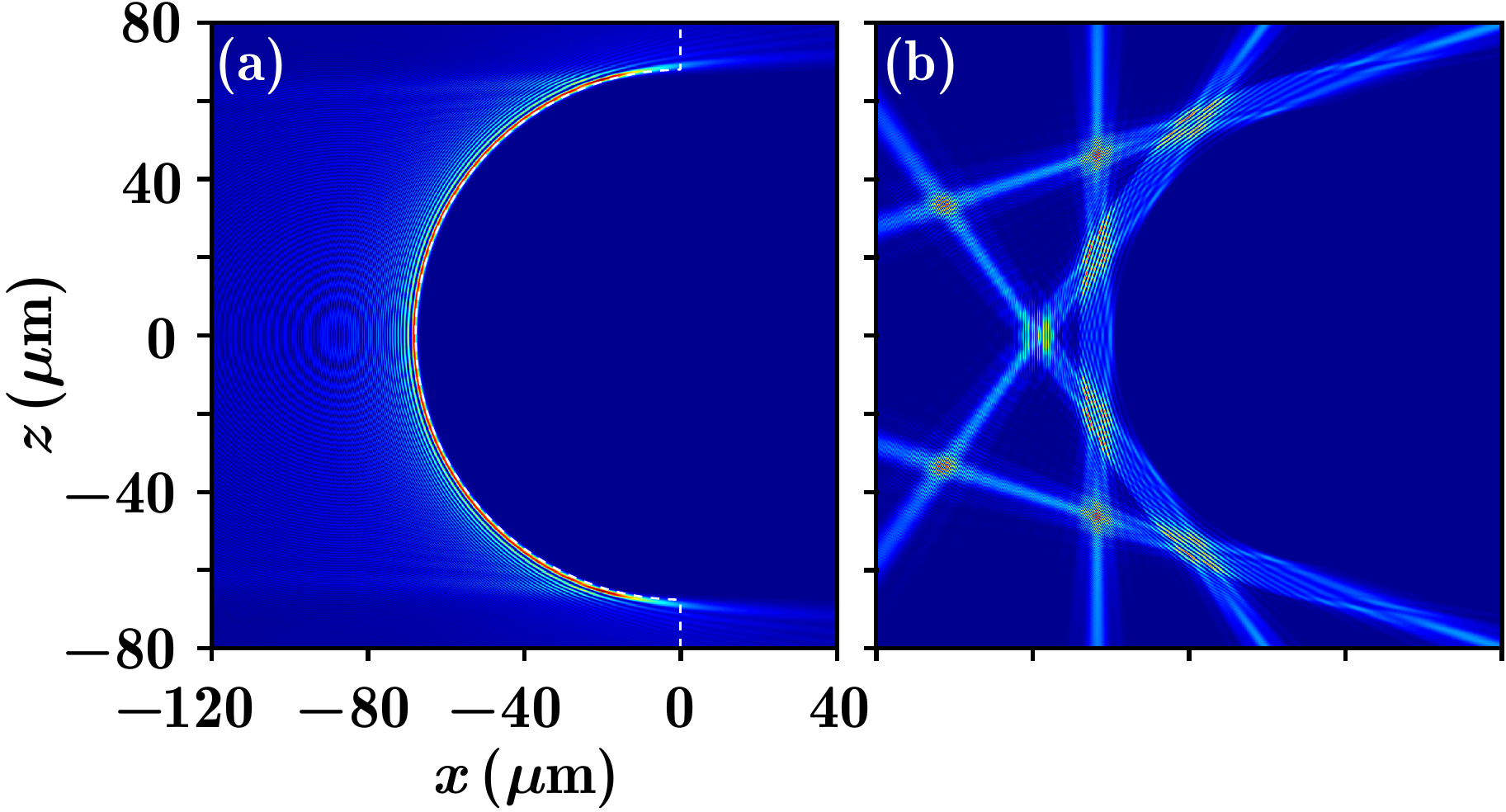}
  \caption{(Color online) (a) Nonparaxial accelerating beam with $m=800$. The dashed curve is the theoretical trajectory.
  (b) Nonparaxial accelerating Talbot effect from the superposition of nonparaxial accelerating beams with $m$
  changing from 700 to 800 and $\Delta m=10$. The mode of presentation is similar to Fig. 3 in \cite{lumer.prl.115.013901.2015}.
  \label{fig1}}
\end{figure}

In Fig. \ref{fig1}(a), we display the shape-preserving solution according to Eq. (\ref{eq1}) with $m=800$,
and the dashed curve indicates the theoretical trajectory.
A superposition of two solutions will form a breather along the semicircular trajectory \cite{kaminer.prl.108.163901.2012}.
One obtains the nonparaxial accelerating Talbot effect when the solution in Eq. (\ref{eq2}) is used,
with many components accelerating in unison,
if the difference in $m$ between two nearest components is equal and fixed, and $c_m \equiv 1$,
as shown in Fig. \ref{fig1}(b), which is similar to Fig. 3(a) in \cite{lumer.prl.115.013901.2015}.

Similar to the nonparaxial accelerating breathers,
the periodicity of the nonparaxial accelerating Talbot effect is
also determined by the difference between the two nearest values of $m$.
In light of the circular trajectory,
it is natural to use the \textit{Talbot angle} instead of the Talbot length, to explore the self-images.
For convenience, one may rewrite Eq. (\ref{eq2}) in polar coordinates $(k,\theta)$ by taking $k_x=k\sin(\theta)$ and $k_z=k\cos(\theta)$
\begin{align}\label{eq3}
  E_y(x,z) = \int_{-\pi/2}^{\pi/2} d\theta   \left\{ \exp\left[ik x \sin(\theta) + i k z \cos(\theta) \right] \sum_{m}  c_m \exp\left( im \theta \right) \right\}.
\end{align}
To find the Talbot angle, one should find an angle that makes the summation in Eq. (\ref{eq3}) not affected by $m$.
To this end, one can pick the two nearest components as $m_0 + n\Delta m$ and $m_0+(n+1)\Delta m$,
with $m_0$ being the reference value of $m$ in Eq. (\ref{eq3}),
$n$ an arbitrary integer, and $\Delta m$ the radial difference between the two nearest components.
Thus, the superposition of the two components is
\begin{align}\label{eq4}
  \exp[ i(m_0+ n\Delta m) \theta ] + \exp\{i [m_0+(n+1)\Delta m] \theta \} = 
  \exp[ i(m_0+ n\Delta m) \theta ] [1+\exp(i\Delta m \theta)].
\end{align}
Clearly, if $\Delta m \theta$ is an integer multiple of $2\pi$, the value of the expression in Eq. (\ref{eq4}) can be rewritten as
$  2\exp ( im_0 \theta )$,
which is independent of $m$.
As a result, one may define the Talbot angle,
\begin{equation}\label{eq5}
  \theta_T = \frac{2\pi}{\Delta m}.
\end{equation}
From Eq. (\ref{eq5}), one finds that the Talbot angle is inversely proportional to the radial difference $\Delta m$.
The smaller $\Delta m$, the larger the Talbot angle.
For the case in Fig. \ref{fig1}(c), the Talbot angle is $\pi/5$.

As reported in Ref. \cite{kaminer.prl.108.163901.2012},
to observe a periodic behavior,
the two solutions should accelerate in unison.
Even though the transverse displacement does not affect the unisonant oscillation of paraxial accelerating beams,
which helps in explaining the paraxial accelerating breathers and Talbot effect \cite{driben.ol.19.5523.2014,lumer.prl.115.013901.2015,zhang.ol.40.5742.2015},
one cannot apply this to the nonparaxial accelerating beams.
The reason is that one has to make sure the sum $\sum_m \exp[ik\sin(\theta)(x+m\Delta x)]$ is independent of $m$,
which demands $\Delta x$, the transverse displacement, to fulfill the relation $\Delta x = 2\pi/[k\sin(\theta)]$,
and this is impossible due to the factor $\sin(\theta)$ in the denominator.
That is, the transversely displaced components cannot accelerate  in unison.
Therefore, we construct the general solution as displayed in Eq. (\ref{eq2}) which is based on different equaldistant values of $m$
that were previously reported in \cite{lumer.prl.115.013901.2015},
even though the solution in Eq. (\ref{eq2}) with arbitrary transverse displacements is also a solution of the Helmholtz equation.

\begin{figure}[htbp!]
  \centering
  \includegraphics[width=0.5\columnwidth]{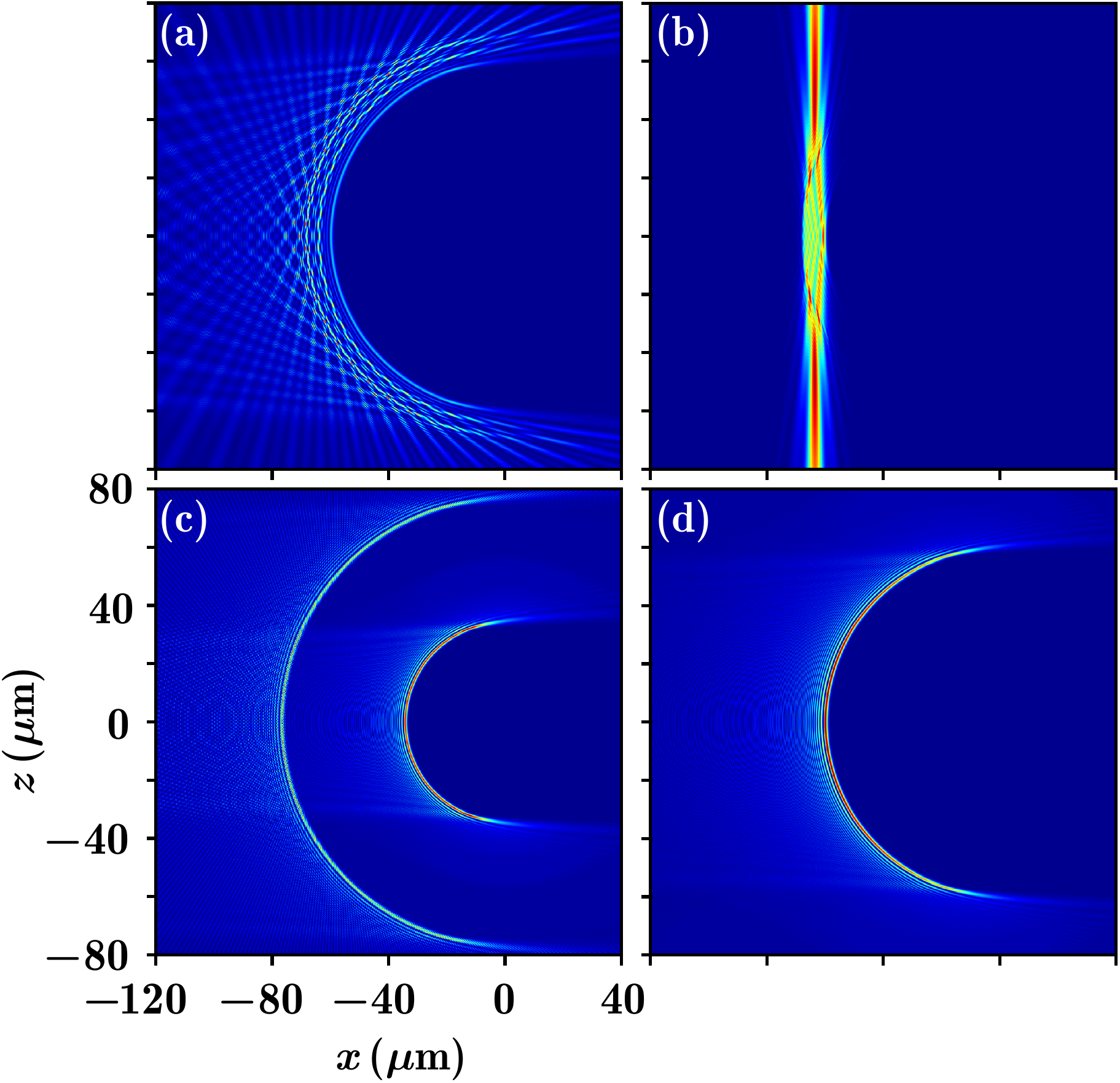}
  \caption{(Color online) Intensity distributions of the superposed nonparaxial accelerating beams.
  (a) $500\le m\le900$ with $\Delta m=50$,
  (b) $700\le m\le716$ with $\Delta m=2$,
  (c) $m=400$ and $m=900$,
  (d) $m=700$ and $m=701$. These beams qualitatively agree with the experimental beam presented in Fig. 3(b) of \cite{lumer.optica.2.886.2015}.}
  \label{fig2}
\end{figure}

From Fig. \ref{fig1}(a), one may observe that the beam deforms when it bends close to $\pi/2$ gradually, especially at the outer rings.
Therefore, the nonparaxial accelerating Talbot effect is getting worse with the increasing of the bending angle.
One can obtain a well-resolved self-imaging in a quite large angle $\sim 2\pi/5$ that is close to $\pi/2$.
Since the Talbot angle is inversely proportional to $\Delta m$,
one may increase the value of $\Delta m$ to obtain a more precise accelerating Talbot carpet with a smaller Talbot angle.
In Figs. \ref{fig2}(a) and \ref{fig2}(b), we exhibit the intensity distributions composed by 9 components
in $500\le m\le900$ with $\Delta m=50$ and in $700\le m\le716$ with $\Delta m=2$, respectively.
One can see that the quality of resolution of the nonparaxial Talbot effect in Fig. \ref{fig2}(a) is much improved.
However, in Fig. \ref{fig2}(b), where the Talbot angle $\theta_T=\pi$,
the superposition of nonparaxial accelerating beams cannot form Talbot effect,
since it propagates along a straight line, without bending. One can also observe that our theoretical beams in Figs. \ref{fig1} and \ref{fig2} generally agree with the experimental curve presented in Fig. 3(b) of \cite{lumer.optica.2.886.2015}.

Going back to the Talbot angle, as expressed in Eq. (\ref{eq5}),
we note an interesting feature by considering two limiting cases $\Delta m\rightarrow 0$ and $\Delta m \rightarrow \infty$,
which correspond to infinite and zero Talbot angles.
We first discuss the $\Delta m\rightarrow 0$ case, which leads to $m_0+n\Delta m\approx m_0$,
so that the radii of the components are almost the same, that is $\sim m_0/k$.
In other words, all the components reduce to one, and one can only see the $m_0$ component, in fact.
From this point of view, a single nonparaxial accelerating beam itself is a case of nonparaxial accelerating Talbot effect with $\theta_T\rightarrow\infty$.
Considering the periodicity of the angle, the maximum Talbot angle is $\pi$.
On the other hand, when $\Delta m\rightarrow \infty$ the radii of the components with $m_0+n\Delta m$ and $n\neq0$ approach infinity.
Therefore, again, one can observe only one component, $m_0$.
Thus, one may also state that a nonparaxial accelerating beam itself is a case of nonparaxial accelerating Talbot effect with $\theta_T\rightarrow0$.
Similar to the paraxial accelerating beams \cite{zhang.ol.40.5742.2015}, the nonparaxial accelerating beams also possess duality.

In Figs. \ref{fig2}(c) and \ref{fig2}(d), we present the intensity distributions of
two superposed nonparaxial accelerating beams, by choosing $(m_0,\Delta m)$ as $(400,500)$ and $(700,1)$, respectively.
In accordance with our prediction, the component with $m_0+\Delta m$
has a much bigger radius than the component with $m_0$, as in Fig. \ref{fig2}(a).
So, the interference between the two components is weakened greatly,
and ultimately only the component with $m_0$ is left with the continuously increasing $\Delta m$.
The situation is opposite in Fig. \ref{fig2}(d) --- the two components are almost the same, due to small $\Delta m$.
As a result, one seemingly finds that there is only one component with $m_0$.
Thus, one can consider the nonparaxial accelerating beam as the image of itself, with the Talbot angle being infinity or zero.

\begin{figure}[htbp!]
  \centering
  \includegraphics[width=0.5\columnwidth]{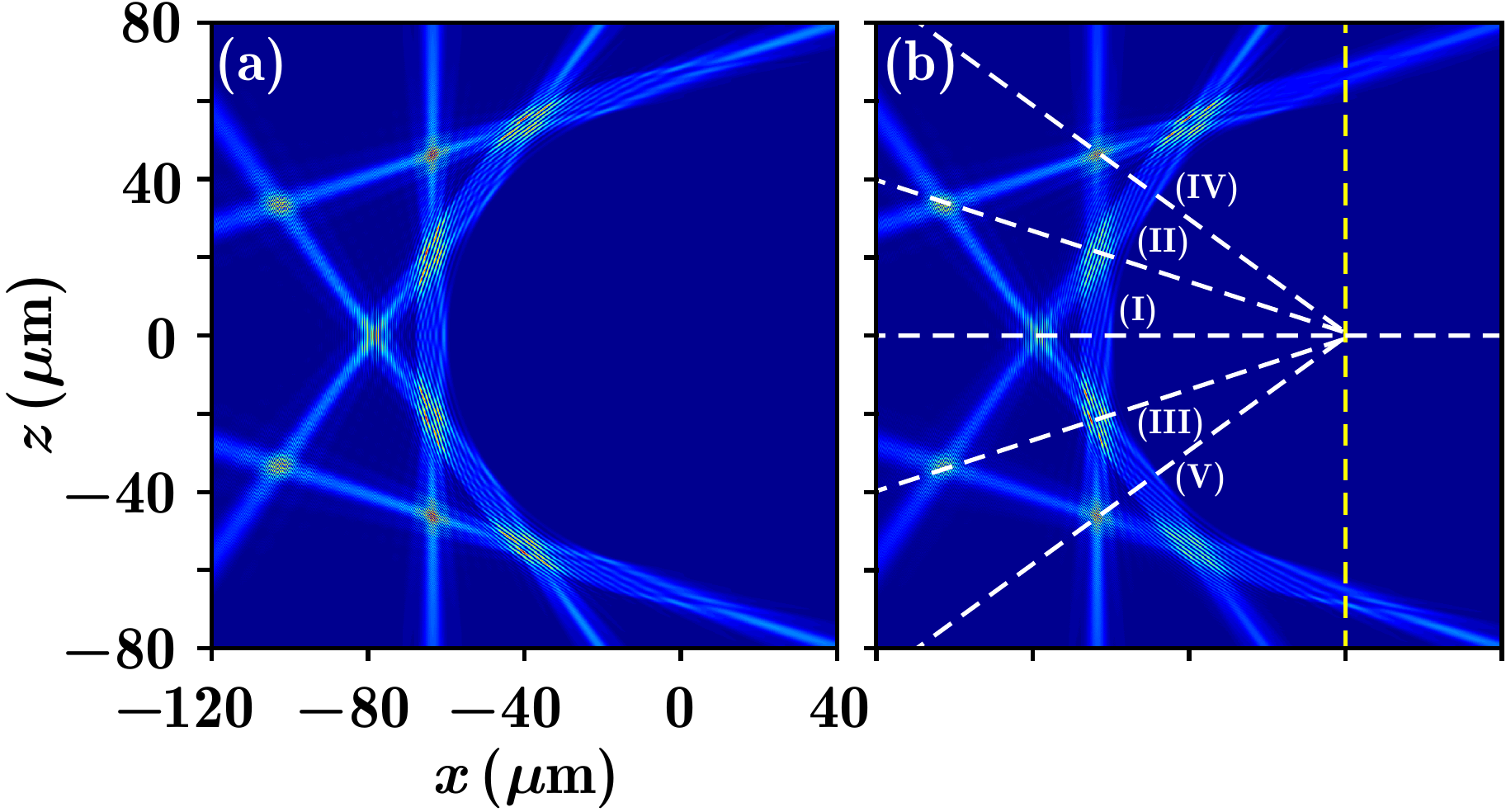}
  \caption{(Color online) Same as Fig. \ref{fig1}(c), but for $\Delta m=5$, and
  (a) $c_m=[\cdots,1,0,1,0,\cdots]$,
  (b) $c_m=[\cdots,1,i,1,i,\cdots]$.}
  \label{fig3}
\end{figure}

As demonstrated in \cite{zhang.ol.40.5742.2015}, the coefficients $c_m$ do not have to be 1 for all the components.
One may choose, e.g., $c_m=[\cdots,1,0,1,0,\cdots]$ and $c_m=[\cdots,1,i,1,i,\cdots]$ to still obtain the Talbot effect.
If we assume that the coefficients for odd components are 0 or $i$, then the summation in Eq. (\ref{eq3}) can be written as
\begin{subequations}\label{eq6}
\begin{equation}\label{eq6a}
\exp(i m_0 \theta)\sum_{n\in\mathbb{Z}} \exp(i2n\Delta m \theta),
\end{equation}
for $c_m=[\cdots,1,0,1,0,\cdots]$, and
\begin{equation}\label{eq6b}
\exp(i m_0 \theta)[1+ i \exp(i\Delta m \theta)] \sum_{n\in\mathbb{Z}} \exp(i2n\Delta m \theta),
\end{equation}
for $c_m=[\cdots,1,i,1,i,\cdots]$.
\end{subequations}
From Eq. (\ref{eq6a}), one can find that the Talbot angle can be written as
\begin{equation}\label{eq7}
  \theta_H=\frac{\pi}{\Delta m},
\end{equation}
which is halved in comparison with Eq. (\ref{eq5}).
That is, the Talbot angle for this case is the same as that in Fig. \ref{fig1}(c), if $\Delta m=5$ is chosen.

\begin{figure}[htbp!]
  \centering
  \includegraphics[width=0.5\columnwidth]{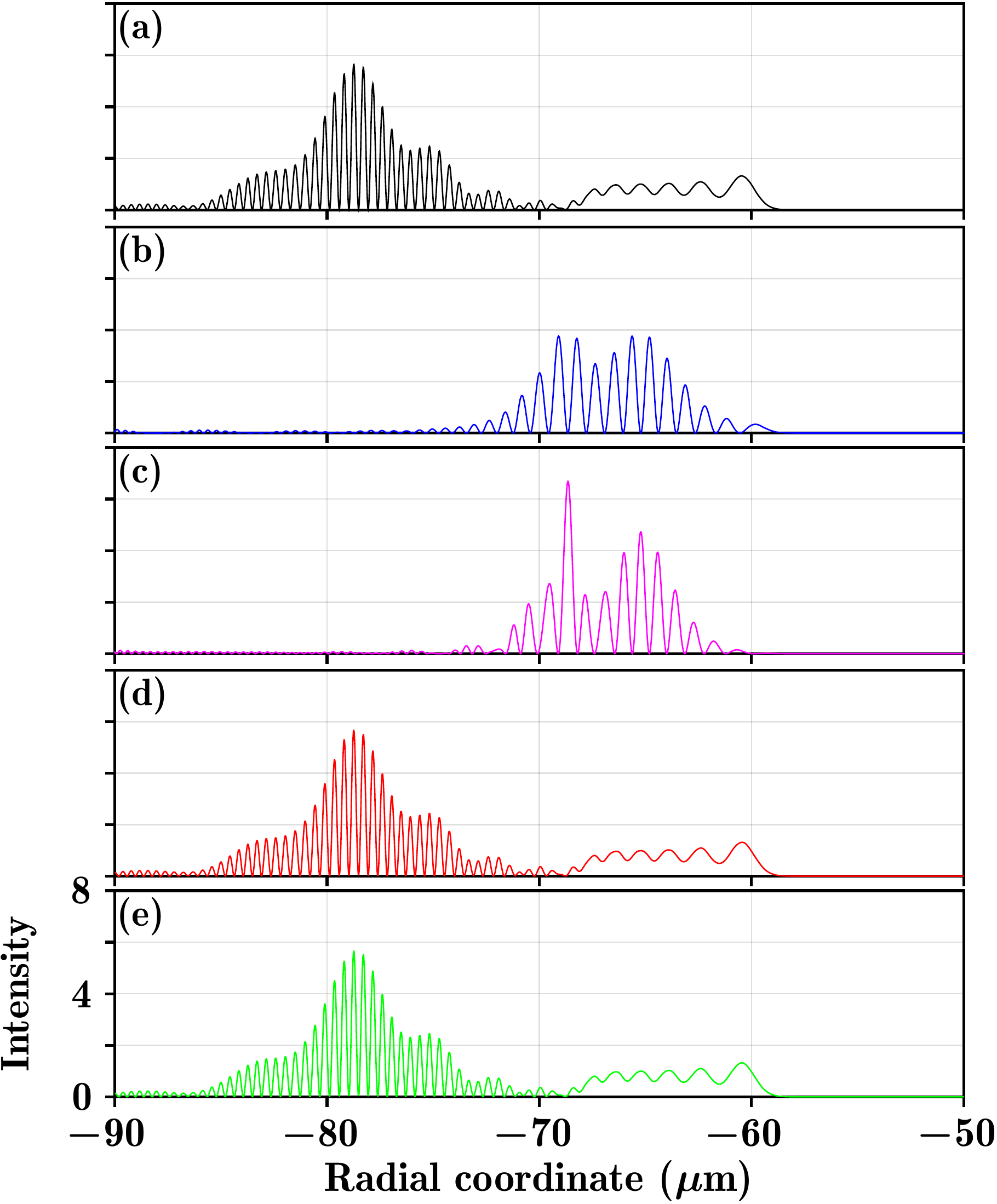}
  \caption{(Color online) (a)-(e) Intensity distributions along the angles marked with (I)-(V) in Fig. \ref{fig3}(b).}
  \label{fig4}
\end{figure}

We display the intensity distribution of superposed nonparaxial accelerating beams in Fig. \ref{fig3}(a),
with the same condition as the one used in Fig. \ref{fig1}(c) except for $\Delta m=5$ and $c_m=[\cdots,1,0,1,0,\cdots]$.
Indeed, the intensity distribution in Fig. \ref{fig3}(a) is same as that in Fig. \ref{fig1}(c).
However, if $c_m=[\cdots,1,i,1,i,\cdots]$ is chosen and other parameters remain the same,
one obtains the intensity distribution as shown in Fig. \ref{fig3}(b).
Even though Figs. \ref{fig3}(a) and \ref{fig3}(b) look the same,
the intensity distributions along the angles are different,
as displayed in Figs. \ref{fig4}(b) and \ref{fig4}(c),
which is not the case in Fig. \ref{fig3}(a).
The reason is that the term $i \exp(i\Delta m \theta)$ in Eq. (\ref{eq6b})
equals $-1$ when $\theta=\theta_H/2$ and 1 when $\theta=-\theta_H/2$,
which indicates that the peaks and valleys of the interference fringes in Fig. \ref{fig4}(b) and Fig. \ref{fig4}(c) are the opposite.
There, the intensity distributions in Figs. \ref{fig4}(a), \ref{fig4}(d) and \ref{fig4}(e) are the same,
however one will find the difference if the phase is also taken into account,
because the term $i \exp(i\Delta m \theta)$ is $i$ in Fig. \ref{fig4}(a), and $-i$ in both Figs. \ref{fig4}(d) and \ref{fig4}(e).
So, the images along the angles marked with (IV) and (V) in Fig. \ref{fig3}(b) are not the Talbot images of (I).
In fact, (IV) and (V) are the mutual Talbot images, because the intensity and phase for both cases are the same.
If one assumes (I) represents the incident beam, then (IV) and (V) are the fractional Talbot images.
For this case, the Talbot angle is still $\theta_T$ instead of $\theta_H$.
We believe that other interesting fractional Talbot images can be obtained when the coefficients of the components are appropriately chosen.

\section{Conclusion}

In summary, we have demonstrated the fractional nonparaxial accelerating Talbot effect among the beams that accelerate along semicircular trajectories,
by choosing the coefficient for each components properly.
The superposed nonparaxial accelerating beams should be concentric, to make the beams accelerate in unison.
We have found that the difference in radius of adjacent beams determines the Talbot angle and
that they display an inverse proportionality.
Similar to the Airy beam, a nonpraxial accelerating beam also is a Talbot effect of itself,
with the Talbot angle being $\pi$ or zero.
We believe that our work not only enriches the Talbot effect family,
but also broadens the practical utility of nonparaxial accelerating beams.

\section*{Acknowledgements}
The National Basic Research Program of China (2012CB921804);
National Natural Science Foundation of China (61308015, 11474228);
Key Scientific and Technological Innovation Team of Shaanxi Province (2014KCT-10);
and Qatar National Research Fund  (NPRP 6-021-1-005).
MRB also acknowledges support by the Al Sraiya Holding Group.

\bibliographystyle{myprx}
\bibliography{my_refs_library}

\end{document}